\documentclass[aps,pra,a4,11pt,preprint]{revtex4-1}
\usepackage[utf8]{inputenc}
\usepackage[T1]{fontenc}
\usepackage{hyperref}
\usepackage{mathpazo}
\usepackage{amsmath,amssymb,amsfonts,bbm,graphicx,times,color,mathptmx,hyperref}
\usepackage{mathrsfs}
\newcommand{\corr}[1]{{\leavevmode  #1}}
\newcommand{\corrbina}[1]{{\leavevmode #1}}
\def\imm{\mathrm{i}}

\begin{document}
\title{Noisy propagation of Gaussian states in optical media with finite bandwidth}
\author{Berihu Teklu}
\author{Matteo Bina}
\author{Matteo G. A. Paris}
\affiliation{Department of Applied Mathematics and Sciences and Center for Cyber-Physical Systems (C2PS), 
Khalifa University, 127788, Abu Dhabi, United Arab Emirates}
\affiliation{Quantum Technology Lab, Dipartimento di Fisica {\em Aldo Pontremoli}, 
Universit\`a degli Studi di Milano, I-20133 Milano, Italy}
\affiliation{INFN - Sezione di Milano, I-20133 Milano, Italy}
%%%%%%%%%%%%%%%%%%%%%%%%%%%%%%%%%
\begin{abstract}
We address propagation and entanglement of Gaussian states in optical media characterised 
by nontrivial spectral densities. In particular, we consider environments with a finite bandwidth $J(\omega) = J_0 
\left[\theta(\omega -\Omega) - \theta (\omega - 
\Omega - \delta)\right]$, and show that in the low temperature regime $T\ll \Omega^{-1}$: i) secular terms in the master equation may be neglected; 
ii) attenuation (damping) is strongly suppressed; iii) the overall 
diffusion process may be described as a Gaussian noise 
channel with variance depending only on the bandwidth.
\corr{We find several regimes where propagation is not much detrimental and entanglement may be protected form decoherence.} 
\end{abstract}
\maketitle
\date{\today}
%%%%%%%%%%%%%%%%%%%%%%%%%%%%%%%%%
\section{Introduction}\label{s:intro}
Engineering and control of quantum systems in the presence of noise is a crucial step in the development of quantum technology \cite{GardinerZoller, Soare, Levy, GenoniPRL,nlnoise,mgap95,Myatt,PiiloManiscalco,Bellomo07,Bellomo13}. In turn, much attention has been devoted to 
describe the dynamics of open quantum systems for different kind 
of environments, i.e. different sources of damping and decoherence \cite{B.Petruccione,Xu,Dehghani,Dehghani20}. 
It is often challenging to obtain the exact dynamics of an open quantum system and different kinds of approaches have been developed to derive an approximate description at different levels of accuracy. 
Besides assuming a weak coupling, the approximations usually employed 
to obtain analytic master (or Langevin) equations \cite{GKS} for the system under 
investigation include neglecting memory effects 
(Markovian approximation), using time coarse graining (not accounting for potential contributions coming from the short-time dynamics), and assuming some simplified
form for the spectral density of the environment, e.g. a Lorentzian one
or an Ohmic one with a large frequency cut-off. 
\par
On the other hand, there are several systems where nontrivial spectral  densities appear, e.g. polarons in metals and semi-conductors, photonic 
crystals and micro-mechanical oscillators, and a question arises on 
the effect of the interaction with this kind of media on the quantum 
properties of a given system \cite{Legget,Shnirman,Paavola,Martinazzo}. 
Moreover, memory effects, backflow of information and short-time dynamics can play a significant role in these scenarios and a non-Markovian approach should be employed \cite{BreuerVacchini,NonMarkov_Rev_Vega,TrapaniBina}. Structured environments have been thoroughly studied and characterized from the point of view of quantum probing, both in the continuous- and discrete-variables regimes \cite{BinaGrasselli,BinaSalari,GebbiaBenedetti}, in the context of bath engineering for controlled systems.
In particular, we focus attention on
composite systems where the propagation of frequencies in a limited range 
is forbidden, owing to the spectral 
properties of their constituents. \corr{A relevant example is that of photonics crystals made of materials with very 
different optical properties (e.g. different refraction indices), overall resulting in the creation of photonic band gaps \cite{PriorPlenio,foresi97, Qiao99,Liu17,Sheng18}.}
\par
In this paper, motivated by some recent experimental implementations in photonic crystal wave guides \cite{Hood, Yu, Javadi}, we address the propagation of Gaussian states of light through a medium characterized 
by a finite bandwidth $\delta$, i.e. the spectral density $J(\omega)$ 
of the structured environment is non-vanishing only in a given interval $[\Omega, \Omega + \delta]$. 
\corr{Though our analysis is based on a simple model, }
it shows that these spectral features result in a peculiar
short-time dynamics of the system, where the secular terms may be neglected, 
pure attenuation is strongly suppressed, and the overall dynamics may 
be described as a Gaussian noise channel with variance depending only 
\corr{on the bandwidth parameters $\delta$ and $J_0$ and not on the natural frequency $\omega_0$ of the 
mode, nor on the location $\Omega$ of the bandwidth in the spectrum.} We then use our results to study the propagation of Gaussian entangled states in media with finite bandwidth.
\par
The paper is stuctured as follows: in Sec. \ref{s:EME}, we describe 
our model and present the non-Markovian master equation. 
Section \ref{s:EDynGS} illustrates the solution of the 
master equation with an initial Gaussian state. In addition, we review how to quantify entaglement for two-mode continuous-variable (CV) 
Gaussian state. 
In Sec. \ref{s:ED} we investigate the validity of the secular 
approximation, and discuss the dynamics of entanglement. 
Sec. \ref{s:concl} closes the paper with some concluding remarks. 
%%%    
\section{The interaction model}  \label{s:EME}
Let us consider a single-mode field at natural frequency $\omega_0$, the system, interacting with an environment that 
we assume at thermal equilibrium.  The interaction Hamiltonian in natural units 
may be written as 
\begin{equation}\label{eq:Hgen}\begin{split}
H&=\frac{\omega_0}{2}\left( P^2 + X^2 \right)+ \sum_{n} \frac{\omega_n}{2} 
\left ( P_n^2+X_n^2 \right) 
%\\ &
-\alpha\, X\otimes \sum_{j} \gamma_j X_j,
\end{split} \end{equation}
where $\alpha$ (dimensionless) is the overall coupling strength 
between the system and the environment, $X$ and $P$ are the canonical 
operators of the system mode, $X_j$ and $P_j$ are operators of the 
environmental modes. We remind that in terms of the field mode, the 
quadrature operators are given by $X(\varphi)=(a\, {
\rm e}^{-\imm\varphi}+a^\dag{\rm e}^{\imm\varphi})/\sqrt{2}$, 
with $X\equiv X(0)$ and $P\equiv X(\pi/2)$. Finally, the quantities 
$\omega_j$ denote the frequencies of the 
environmental modes, and  $\gamma_j$ are the (dimensional) couplings 
between the system and the $j$-th environmental mode.  
At $t=0$ we assume a factorized state $\varrho_0\otimes \mathcal{E}$, where $\varrho_0$ is the initial state of the system and $\mathcal{E}$ \corr{an equilibrium (thermal) 
state of the environment, i.e. $\mathcal{E}={\rm e}^{-\beta H_B}/\mathcal{Z}$, 
where $\beta=T^{-1}$ is the inverse temperature}, $H_B$ the free energy of the environment and $\mathcal{Z}$ the partition function. \corr{This is a Gaussian state too}. We use natural
units and thus besides $\hbar=1$ we also have the Boltzmann constant $k_B=1$.
\par
Upon evolving the overall system and tracing out the environmental degrees of freedom we 
obtain a time-local master equation, which describes the noisy evolution
of the system mode \cite{Vasile2009, HuPaz,B.Petruccione}
\begin{equation}\label{eq:ME}
\dot \varrho(t) =  -\imm\big [ H_0,\varrho(t)\big ]+\imm\, r(t)\big[ X^2,\varrho(t)\big]  
-\imm\, \gamma(t)\big[X,\{ P,\varrho(t)\}\big ] -\Delta(t)\big[X,\left[X,\varrho(t)\right]\big] +\Pi (t)\big[X,\left[P,\varrho(t) \right]\big] \,,
\end{equation}
 where $H_0$ is the free Hamiltonian (first term in Eq.~(\ref{eq:Hgen})), and $[\cdot\,,\cdot]$ and 
 $\{\cdot\,,\cdot\}$ denote commutators and anticommutators, respectively. 
 The first term in Eq.~({\ref{eq:ME}}) is due to the unitary part of the  time evolution, whereas
 the second one induces a time-dependent energy-shift. The
 third term is a damping term and the last two are responsible for 
 diffusion. The different time-dependent coefficients link the 
 non-Markovian features of the dynamics with the spectral
 structure of the environment and its thermal excitations.
 Up to second order in $\alpha$ we have
\begin{align}
\label{eq:CoefficientsME}
       \gamma (\tau)&=\int_{0}^{\tau}\!\!\! ds \sin(\omega_0\,
      s)\int_{0}^{\infty}\!\!\!\! d\omega\, J(\omega) \sin(\omega s)\,,\quad
         \Delta (\tau)=
     \int_{0}^{\tau}\!\!\! ds  \cos(\omega_0\, s) \int_{0}^{\infty}\!\!\!\!
        d\omega \coth \frac{\beta \omega}{2} J(\omega)\cos(\omega s)\,, 
         \\
          r (\tau)&= \int_{0}^{\tau}\!\!\! ds \cos(\omega_0\, s)\int_{0}^{\infty}\!\!\!\!d\omega\, J(\omega) \sin(\omega s)\,,  \quad
    \Pi (\tau) =
        \int_{0}^{\tau}\!\!\! ds \sin(\omega_0\, s) \int_{0}^{\infty}\!\!\!\!
        d\omega  \coth \frac{\beta \omega}{2} J(\omega) \cos(\omega s) \,,
\end{align} 
where $J(\omega)=\alpha^2\sum_j \frac{\gamma_j}{2}\delta(\omega-\omega_j)$ 
is the spectral density of the environment. 
\corr{
The average number of thermal excitations for the mode at frequency 
$\omega$ is given by $N(\omega)=\big ({\rm e}^{\beta\omega}-1\big )^{-1} = \frac12 (\coth \frac{\beta \omega}{2}-1)$, i.e. $\coth \frac{\beta \omega}{2} = 2 N(\omega)+1$.}
\par
A general solution of the master equation (\ref{eq:ME}) can be found through the quantum characteristic approach \cite{Intra2003} 
in terms of the canonical variables $\vec{z}=(x\, ,\, p)$, assuming a weak coupling regime which corresponds to fulfill the condition $\alpha\ll 1$:  
\begin{equation}
  \chi [\vec{z}\,](t)=
   {\rm e}^{-\vec{z}^{\,T}\overline{W}(t)\vec{z}}
      \chi\left[e^{-\frac{\Gamma (t)}{2}}R^{-1}(t)\vec{z}\, \right] (0) \, ,
      \label{eq:QCFt}
\end{equation}
where 
\begin{align}
 \Gamma (t) &= 2\int_{0}^{t} \gamma(\tau) d\tau  \label{Gamma} \\
  R(t) &\simeq
   \left( \begin{array}{cc}
      \cos(\omega_0 t) & \sin(\omega_0 t) \\  -\sin(\omega_0 t) & 
       \cos(\omega_0 t)
    \end{array}
   \right)  \qquad M(\tau) =\left( \begin{array}{cc}
    \Delta(\tau) &-\frac{\Pi (\tau)}{2} \\  -\frac{\Pi (\tau)}{2}   & 0 \end{array} \right) \\
 \overline{W}(t)&={\rm e}^{-\Gamma(t)}\big [ R^{-1}(t) \big ]^T W(t) R^{-1}(t) \qquad
 W(t) =\int_0^t{\rm e}^{\Gamma(\tau)}R^T(\tau)M(\tau)R(\tau)d\tau \,.
\end{align}
\corr{Since the Hamiltonian (\ref{eq:Hgen}) is at most bilinear in the system quadrature operators, it is easy to prove that \corrbina{it} induces 
a Gaussian evolution map, i.e. a map which preserves the Gaussian character of any initial Gaussian state \cite{SerafiniB}. For this reason, the resulting dynamics is usually referred to as a Gaussian channel.}
%%%%%%%%%%%%%%%%%%%%%%%%%%%%%%%%%%%%
\section{Dynamical evolution of Gaussian states}\label{s:EDynGS}
In this Section we review the solution of the master equation for Gaussian 
states \cite{Intra2003}. 
In particular, we focus on two-mode Gaussian states (each one interacting with 
its own environment) in order to analyze the dynamics of entanglement. On the other 
hand, the conclusions about the features of the channel are general, and apply to signals with any 
number of modes.
\par
Let us thus consider a single two-mode Gaussian state, with characteristic function at time $t=0$ 
$\chi_0(\vec{z})=\exp\{-\frac{1}{2}\vec{z}^{T}
\sigma_0\,\vec{z}-i\,\vec{z}^{T}\bar{\mathbf{X}}_{in}\}$. The initial  covariance matrix $\sigma_0$  is a $4\times 4$ matrix
\begin{equation}\label{eq:CovMatzero}
\sigma_0=\left(
          \begin{array}{cc}
            \mathbf{A_0} & \mathbf{C_0} \\
            \mathbf{C^T_0} & \mathbf{B_0} \\
          \end{array}
        \right),
\end{equation}
where $\mathbf{A_0}= a\,{\mathbbm I}$, $\mathbf{B_0}=b\,{\mathbbm I}$,
$\mathbf{C_0}={\rm Diag}(c_1,c_2)$, with $a$,$b>0$ and $c_1$, $c_2$
real numbers, and $\mathbbm I$ the $2\times 2$ identity matrix.
 We remind that the system-environment interaction, and thus the time 
evolution, maintains the Gaussian 
character \cite{Stefano05,Vasile2009,Vasile2010,Berihu22}. The evolved state 
is a two-mode Gaussian state with mean and covariance matrix 
 described by 
\begin{align}
\bar{\mathbf{X}}_t&=e^{-\Gamma(t)/2}\big [ R(t)\oplus R(t)\big ]
\bar{\mathbf{X}}_{in} \label{e1} \\
\label{CovMatT}
\sigma_t&=e^{-\Gamma(t)}\big [ R(t)\oplus R(t)\big ] \sigma_0 \big [  R(t)\oplus
R(t)\big ]^T+2(\bar{W}(t)\oplus\bar{W} (t)) \, .
\end{align}
Upon substituting the expression of the coefficients obtained in the weak-coupling approximation, into Eqs.~(\ref{e1}) and (\ref{CovMatT}), 
the covariance matrix at time $t$ may be written as
\begin{equation}\label{CovMatT2}
{\boldsymbol \sigma_{t}}= \left(
\begin{array}{c | c}
\mathbb{\bf A_{t}} &\mathbb{\bf C_{t}} \\ \hline\mathbb{\bf C_{t}}^{T} & \mathbb{\bf A_{t}}
\end{array}
\right),
\end{equation}
where
\begin{align}\label{At}
\mathbb{\bf A_{t}} ={} \mathbb{\bf A_{0}} e^{-\Gamma (t)}
+ \left(
\begin{array}{cc}
\Delta_{\Gamma}(t) + \big [\Delta_{\rm co}(t) - \Pi_{\rm si}(t)\big ] &-\big [ \Delta_{\rm si}(t) - \Pi_{\rm co}(t)\big ] \\
-\big [\Delta_{\rm si}(t) - \Pi_{\rm co}(t)\big ] & \Delta_{\Gamma}(t) - \big [ \Delta_{\rm co}(t) - \Pi_{\rm si}(t) \big ]
\end{array}
\right)
\end{align}
and
\begin{align}\label{Ct}
\mathbb{\bf C_{t}} = \left(
\begin{array}{cc}
c\, e^{-\Gamma (t)}\, \cos(2\omega_0 t) &
c\, e^{-\Gamma (t)}\, \sin(2\omega_0 t) \\
c\, e^{-\Gamma (t)}\, \sin(2\omega_0 t) & -c\, e^{-\Gamma (t)}\,
\cos(2\omega_0 t)
\end{array}
\right).
\end{align}
In order to obtain the compact forms (\ref{At}) and (\ref{Ct}), we have introduced the following expression
\begin{equation}
\Delta_{\Gamma}(t)=e^{-\Gamma(t)}\int_0^t e^{\Gamma(s)}\Delta(s)ds
\end{equation}
and the secular coefficients 
  \begin{subequations}
\label{eq:SecCoeff}
\begin{align}
&\Delta_{co}(t)=e^{-\Gamma(t)}\int_0^t
e^{\Gamma(s)}\Delta(s)\cos[2\omega_0(t-s)]ds \quad \Delta_{si}(t)=e^{-\Gamma(t)}\int_0^t e^{\Gamma(s)}\Delta(s)\sin[2\omega_0(t-s)]ds\\
&\Pi_{co}(t)=e^{-\Gamma(t)}\int_0^t
e^{\Gamma(s)}\Pi(s)\cos[2\omega_0(t-s)]ds \quad
\Pi_{si}(t)=e^{-\Gamma(t)}\int_0^t
e^{\Gamma(s)}\Pi(s)\sin[2\omega_0 (t-s)]ds.
\end{align}\end{subequations}
In situations where the secular terms may be neglected (as we will see, this 
is our case) the diagonal blocks of the covariance matrix evolve as 
\begin{align}
\mathbb{\bf A_{t}} ={} \mathbb{\bf A_{0}}\, e^{-\Gamma (t)}
+  \Delta_{\Gamma} (t)\, {\mathbbm I} \,. \label{e3}
\end{align}
\par
Let us now introduce the main ingredient of our analysis, i.e. a specification of 
the properties of the bosonic reservoirs through the form of the spectral density. 
In order to describe the presence of a finite bandwidth, i.e. the fact that the 
propagation of certain frequencies, or a certain range of frequencies, is forbidden,
we consider a spectral density of the form
$$J(\omega) =J_0 
\left[\theta(\omega -\Omega) - \theta (\omega - 
\Omega - \delta)\right]\,.$$ 
The parameter $\delta$ represents the bandwidth of 
the distribution, $\Omega$ specifies its location within the spectrum and \corrbina{$J_0$ is the amplitude of the spectral density}.
In the low temperature regime, i.e. $T\ll \Omega^{-1}$, we may safely assume
$N(\omega)\approx 0$ and $\coth\frac{\omega}{2T}\approx 1$ 
in Eq. (\ref{eq:CoefficientsME}), and obtain an analytic expression for the
relevant coefficients. \corr{In particular, we get a simplified form of the master equation coefficients
\begin{align}
 \Delta (t)&=J_0\, \delta\, t\,, \qquad
 \gamma(t)=\frac13\, J_0\, \delta\, \omega_{0}\,\Omega\, t^3 \,.
\end{align}
Using these expressions, it is straightforward to obtain the time-integrated functions. We have 
$ \Gamma (t)=\frac16\,J_0\, \delta\, \omega_{0}\,\Omega\, t^4$, and $
\Delta_{\Gamma} (t)=\frac12\, J_0 \delta\, t^2$.
In the following, we express all the quantities in units of $\omega_0$, i.e. we make the replacements $t\rightarrow \tau = \omega_0\,t$, $J_0\rightarrow J_0/\omega_0$, $\Omega\rightarrow \Omega/\omega_0$, and
$\delta\rightarrow \delta/\omega_0$. In this way, we may write the time-integrated functions in terms of the \corrbina{dimensionless} time $\tau$ and the \corrbina{dimensionless} parameters $\delta$, $J_0$, and $\Omega$
\begin{align} 
\Gamma (\tau)=\frac16\,J_0\, \delta\,\Omega\, \tau^4\,, \qquad
\Delta_{\Gamma} (\tau)=\frac12\, J_0\, \delta\, \tau^2\,.
\end{align}
Notice that the damping term depends on all the parameters, whereas the diffusion one does not depend on $\Omega$, i.e. on the location of the bandwidth.
Since the memory effects due to the non-Markovian nature of the environment are taking place over short times, the above results imply that attenuation 
(damping) is strongly suppressed in the propagation through media with a finite 
bandwidth. Moreover, if the secular terms may be neglected, the overall diffusion process 
is governed by Eq. (\ref{e3}) and by the espression of $\Delta_{\Gamma} (\tau)$. 
In other words, in this case the dynamics corresponds to a Gaussian channel where gain and loss compensate each other and the resulting effect is that of a diffusion \cite{Stefano05}. 
This situation is usually referred to as a {\em Gaussian noise channel}. In our case,  the variance of the Gaussian noise is depending only on the bandwidth 
parameters $\delta$ and $J_0$,  
and not on the location $\Omega$.}
%%%%
\par
\subsection{Quantification of entanglement} 
\corr{The entanglement of a two-mode CV system may be quantified by different entanglement monotones, including entanglement negativity \cite{EN} and entanglement of formation \cite{EoF,G:03,EoF2008}. Both may be computed starting from the covariance matrix of the system. Here, for analytic convenience, we adopt the negativity $E_N$ as an entanglement quantifier.}

\corr{\begin{equation}\label{EN}
E_N = \max \left(0, - 2 \log \kappa \right)
\end{equation}
where $\kappa$ is the minimum
symplectic eigenvalue of the partially transposed Gaussian state under investigation.
In terms of the the covariance matrix 
${\boldsymbol \sigma_{t}}$ in Eq.~\eqref{CovMatT2}, 
and exploiting the fact that we are dealing with symmetric states we may write $\kappa$ as 
\begin{align}
\kappa = \sqrt{2}\sqrt{I_1 -I_3-\sqrt{(I_1 -I_3)^2-I_4}}\,.
\end{align}
}
The quantities $I_k$ are the symplectic invariants of the covariance matrix, i.e.
$I_1 = \det[{\bf A_{t}}]$, $I_3 = \det[{\bf C_{t}}]$ and $I_4 = \det[{\boldsymbol \sigma_{t}}]$.
%%%%%%%%%%%%%%%%%%%%%%%%%%%%%%%%%%%%
\section{Validity of the secular approximation and entanglement dynamics}\label{s:ED}
Protecting entanglement during evolution and avoiding entangled sudden death \cite{YuLett, YuScience,Almeida,Xu10,Bellomo07,Bellomo08,Maniscalco08,PraBec2004,Alicki,cazza13} is a major task in different 
areas of quantum information science. In this framework, it has been shown 
\cite{Bellomo} that some beneficial effects may be obtained by engineering 
structured environment such as a photonic bandgap materials
\cite{John90,John94}, and this motivates us to analyze in some details 
the dependence of the dynamics on the parameters of spectral density, and their 
interplay with the natural frequency of the involved modes. 
\par
The evolution induced by the master equation Eq. (\ref{eq:ME}) corresponds to a Gaussian map, i.e., an initial Gaussian state maintains its character. It is thus possible to obtain the expression of the covariance matrix at time and then evaluate entanglement at any time for the two modes initially excited in any entangled Gaussian state \cite{Vasile11}. 
To be specific, we address situations where the system of two modes is initially prepared in a twin-beam (TWB) state, i.e. a maximally entangled Gaussian state having covariance matrix coefficients
 $a=b=\cosh 2r$, $c= \sinh 2 r$ with $r>0$ (see Eq. (\ref{eq:CovMatzero})) and analyze the state propagation in a reservoir  with a finite bandwidth spectral density. 
%%%
\begin{figure}[h!]
\center
\includegraphics[width=0.3\columnwidth]{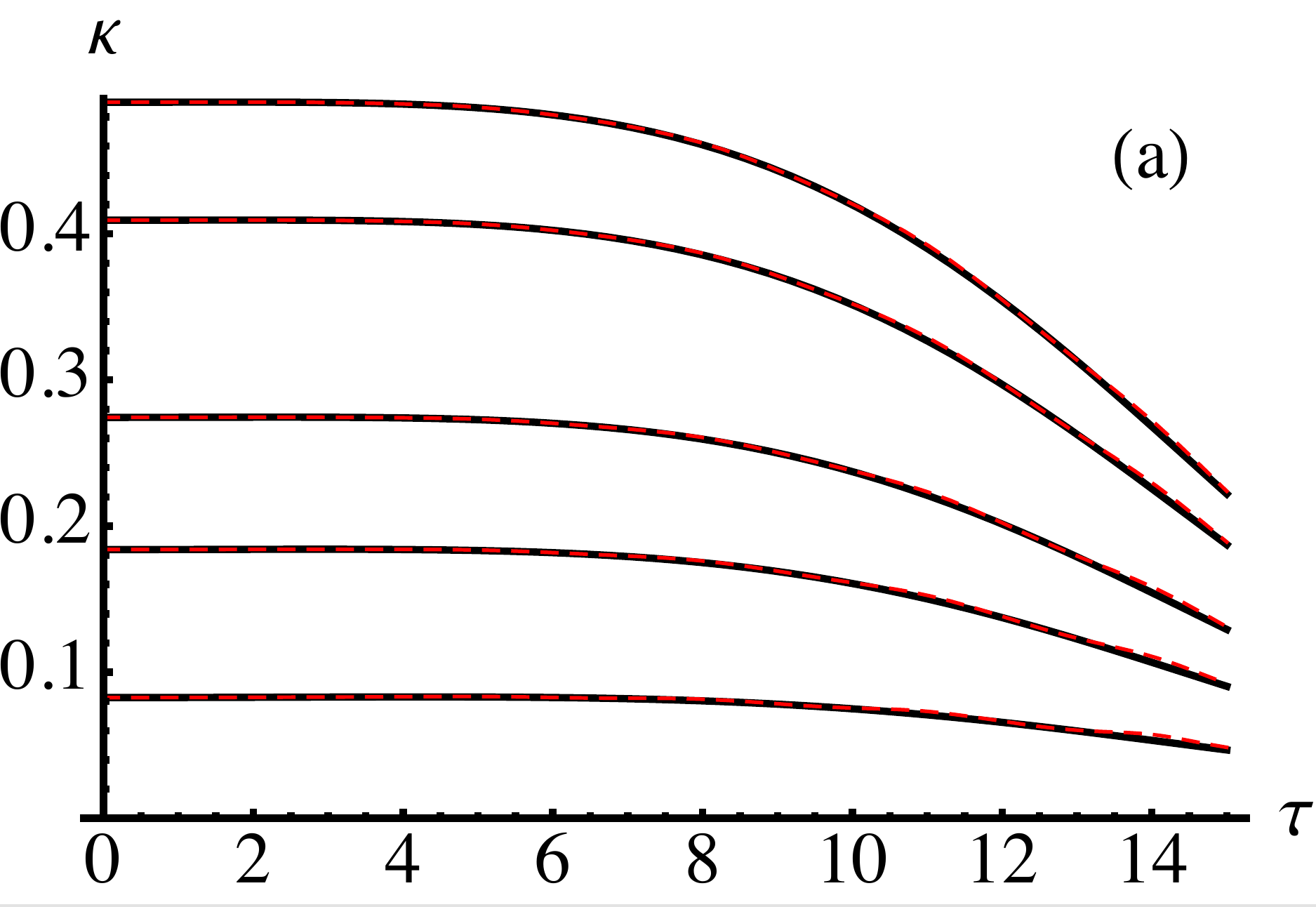}
\includegraphics[width=0.3\columnwidth]{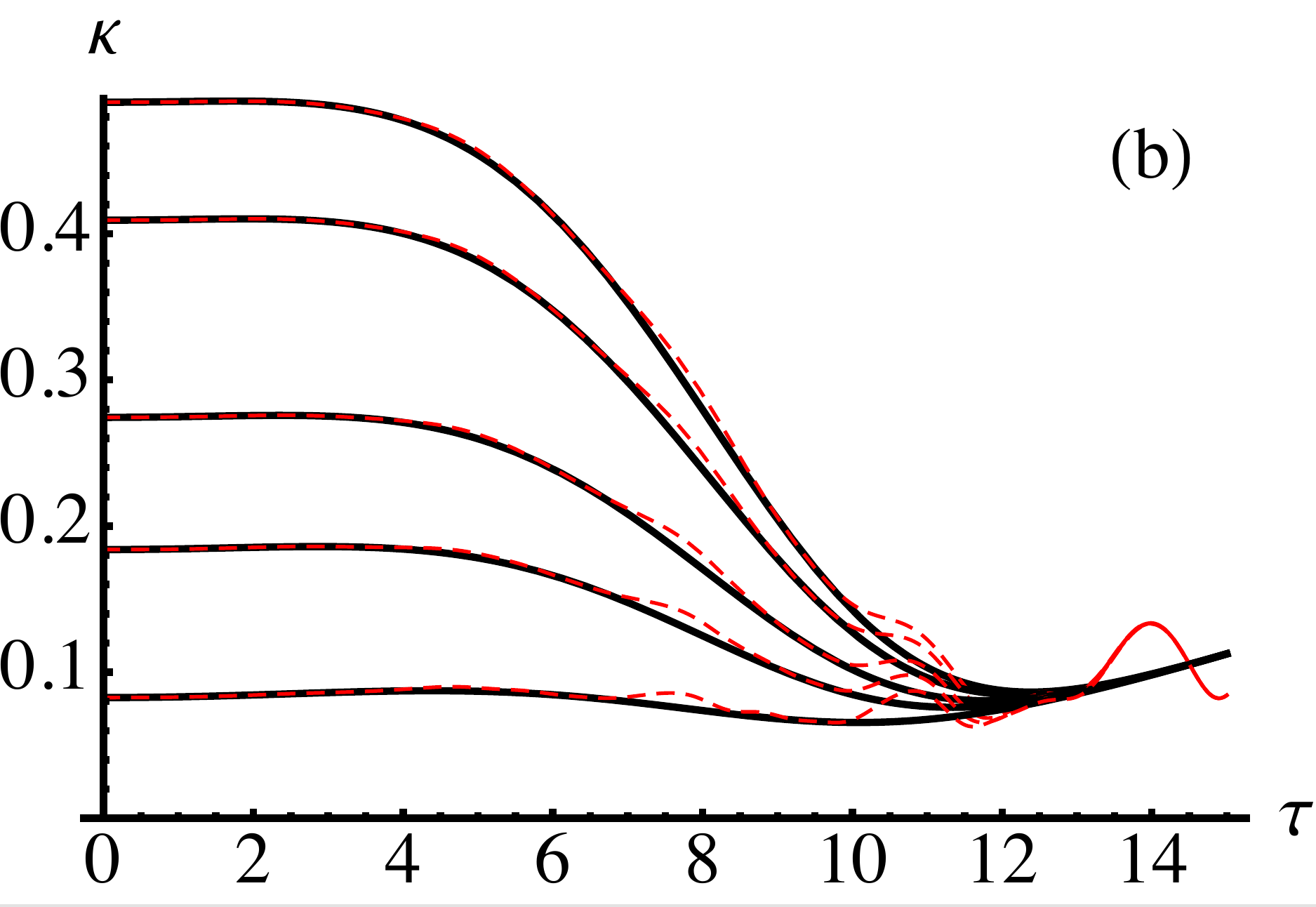}
\includegraphics[width=0.3\columnwidth]{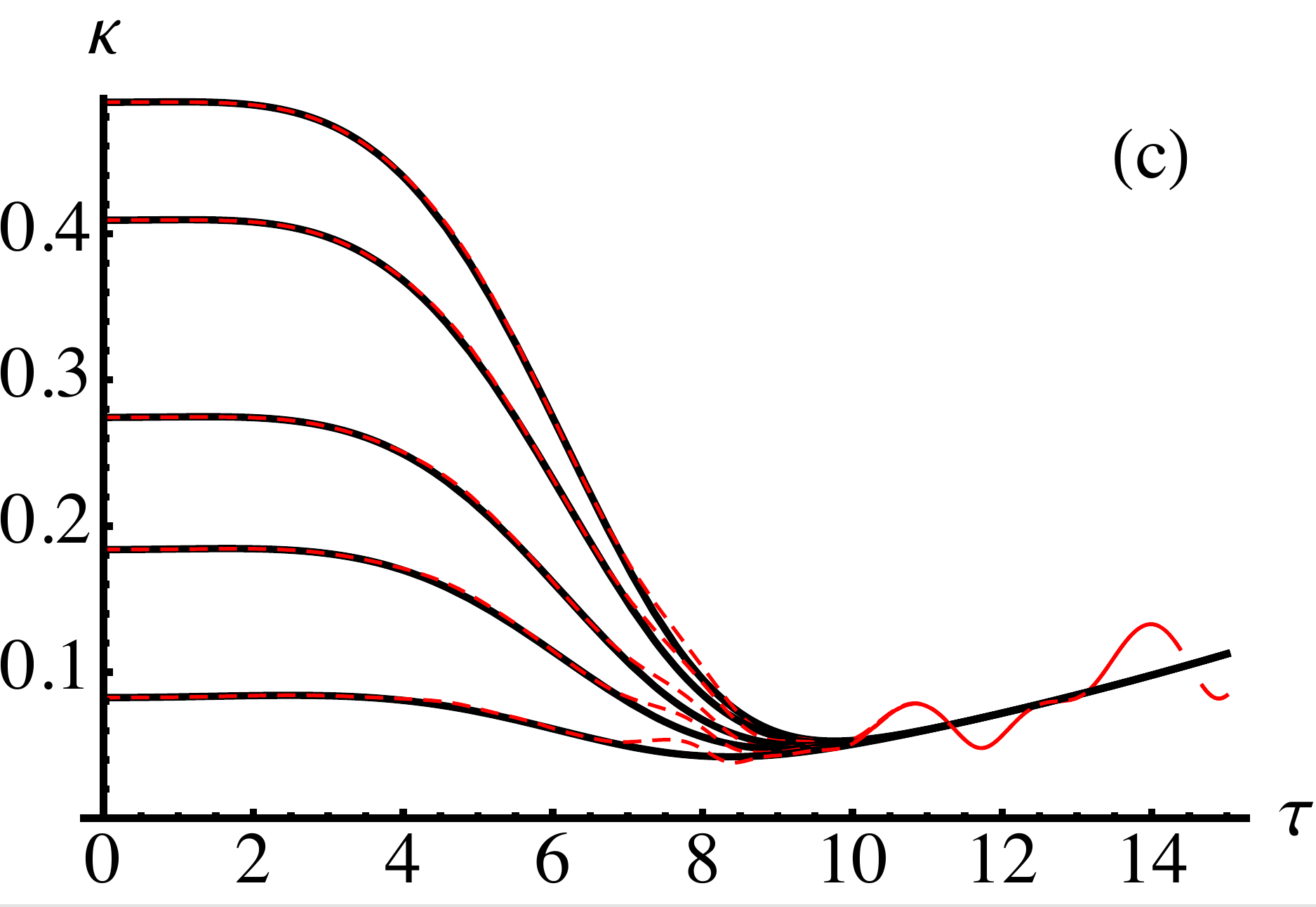}
\caption{\corr{Comparison between the results obtained with and without the secular terms. The three panels show
the symplectic eigenvalue $\kappa$ as a function of the \corrbina{dimensionless} time $\tau$ for different values of the other \corrbina{(dimensionless)} parameters. The solid black curves denote results obtained with the secular approximation (i.e. dropping the secular terms) whereas the red dashed curves denote the full expression including the secular terms. 
In the left panel [panel (a)] we show results 
for $\delta = 10^{-4} $, $\Omega=J_0=1$ and different values of the TWB parameter $r$ (from top to bottom $r=0.01, 0.1, 0.3, 0.5$, and $0.9$). 
In panel (b), we show results for $\delta = 10^{-3} $ and the same values of the other parameters as in panel (a).  In panel (c), we show results for $\delta = 10^{-3} $, $J_0=1$, 
$\Omega=3$ and for different values of the TWB parameter [as in panel (a)].}
\label{f1}}
\end{figure} 
\par
\corr{
At first, we investigate the validity of the secular approximation, i.e. we check when the secular terms may be dropped. To this aim, we calculate the symplectic eigenvalue $\kappa$ with and without the secular terms 
for an initial TWB state with parameter $r$ and compare results for different values of the involved parameters. Using the secular approximation, the expression of $\kappa$ reads as follows
\begin{align}\label{kappa}
\kappa = \frac12 \left( \tau^2\, J_0 \delta + e^{-2 r - \frac16 \tau^4\, J_0 \delta\, \Omega}
\right)
\end{align}
whereas the full expression including the secular terms is rather cumbersome and it is not reported here. Notice also that $\kappa$ depends on the product $J_0 \delta$ rather than on the two parameters independently. We will thus set $J_0=1$ (which means $J_0=\omega_0$ \corrbina{in the original dimensional system}) in the following.
\par
The comparison between the secular and non secular solution is illustrated in Fig. \ref{f1}.
More specifically, in the left panel of Fig. \ref{f1} [panel (a)] we show $\kappa$ as a function of $\tau$ for $\delta = 10^{-4} $, $\Omega=J_0=1$ and for different values of the TWB parameter $r$. In the center panel, $\kappa$ as a function of $\tau$ is shown for $\delta = 10^{-3}$ and the same values of the other parameters. Finally, in the right panel of Fig. \ref{f1} [panel (c)] we show $\kappa$ as a function of $\tau$ for $\delta = 10^{-3}$, $J_0=1$, 
$\Omega=3$ and for different values of the TWB parameter $r$. In all the plots, the solid black lines denote the results without the secular terms and the red dashed lines with the secular terms. \corrbina{We remind that all the involved parameters are in units of $\omega_0$.}

The first observation is that the validity of the secular approximation is a property of the channel, i.e. it is almost independent on the TWB parameter. Moreover, we see that for short times the secular approximation is always valid, independently on the other parameters. How "short" should be the time depends instead on the property of the environment. More precisely, the secular approximation holds for longer times if the bandwidth $\delta$ and the location $\Omega$ are smaller. We also notice that the region of validity of the secular approximation coincides with the region where the dynamics for different values of $r$ differ: when the function $\kappa(\tau)$ no longer depends on the initial value of $r$ the secular approximation starts to fail (though the behaviour of the different functions may be closer also before that time).}

\begin{figure}[h!]
\center
\includegraphics[width=0.3\columnwidth]{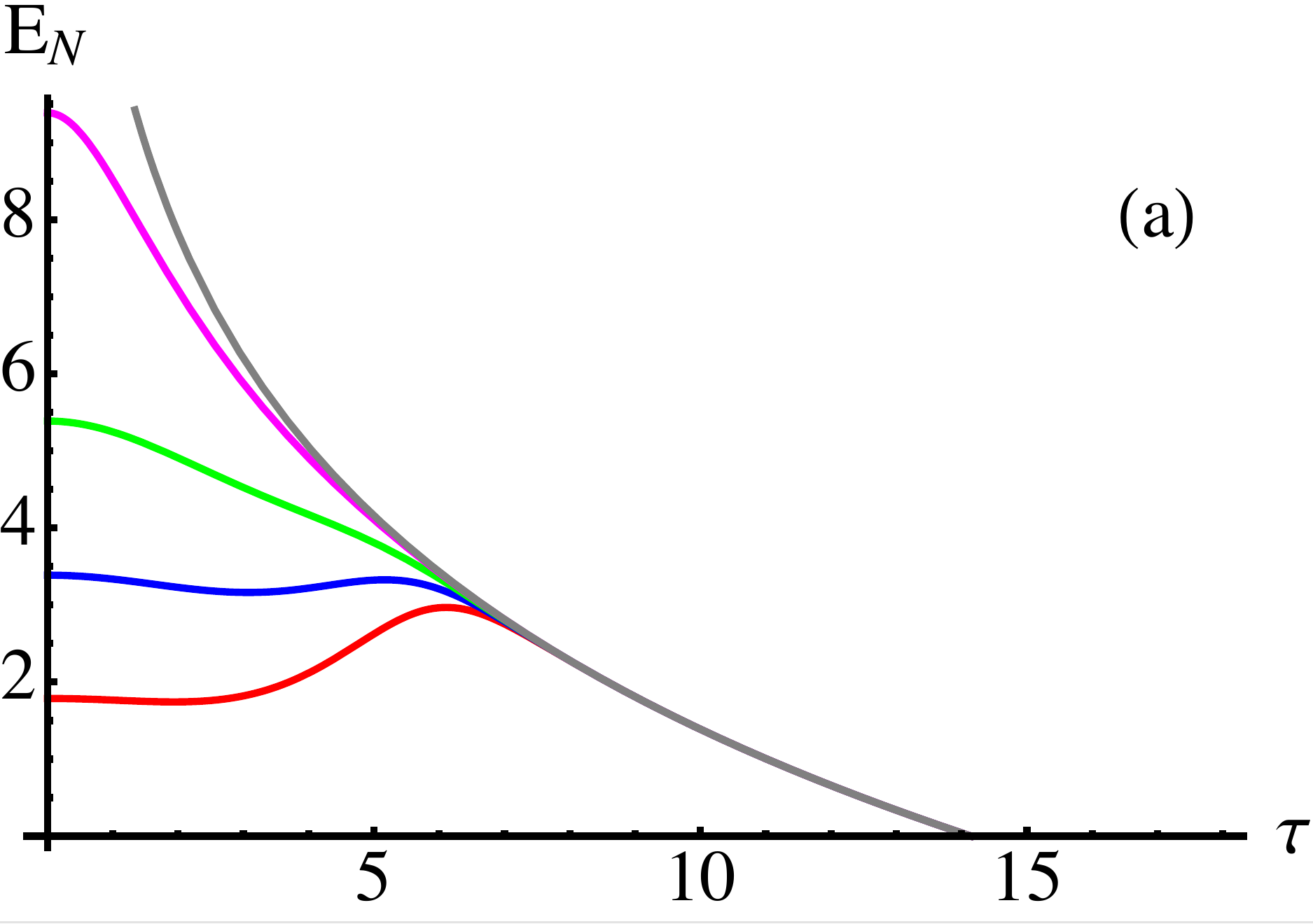}
\includegraphics[width=0.3\columnwidth]{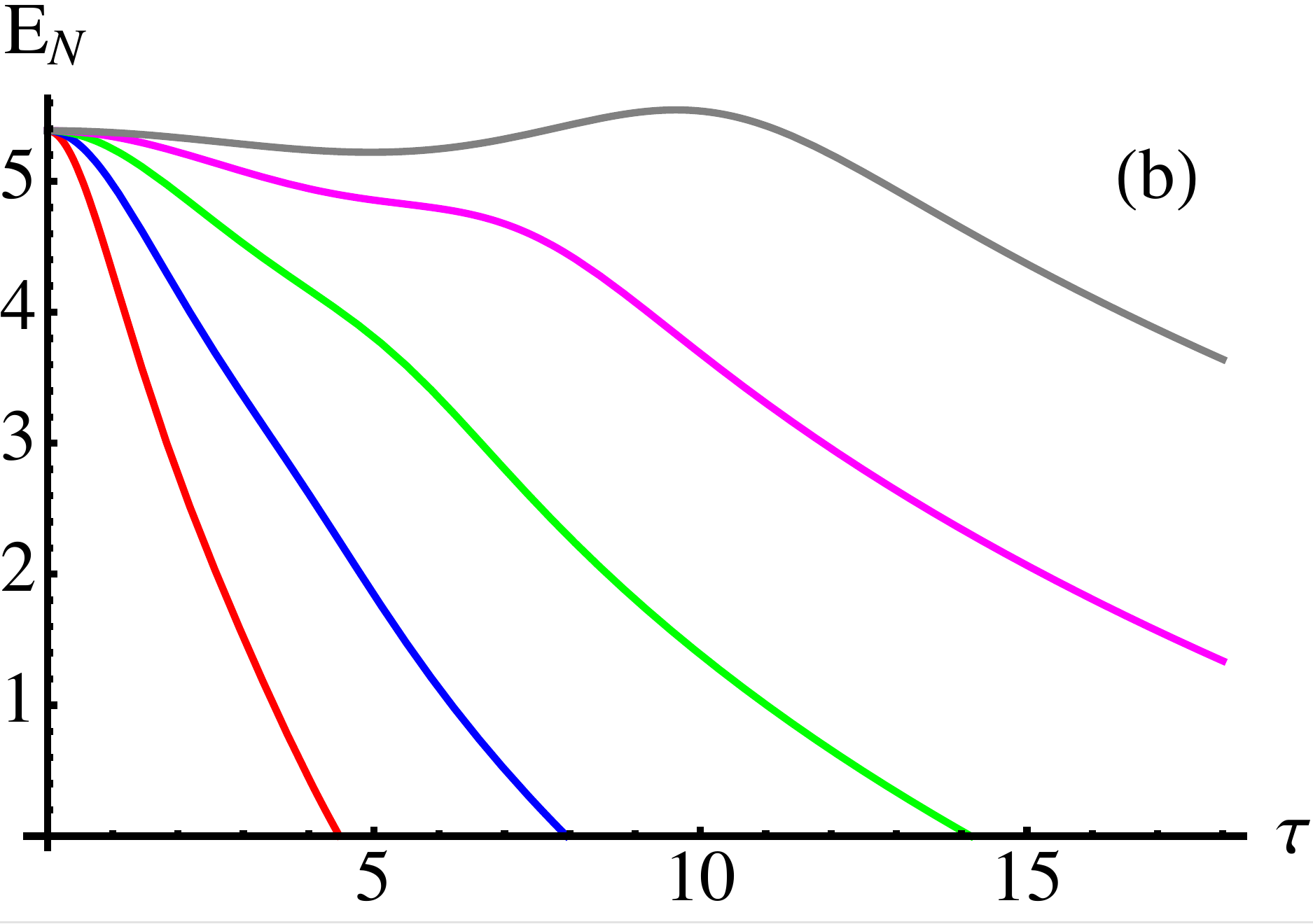}
\includegraphics[width=0.3\columnwidth]{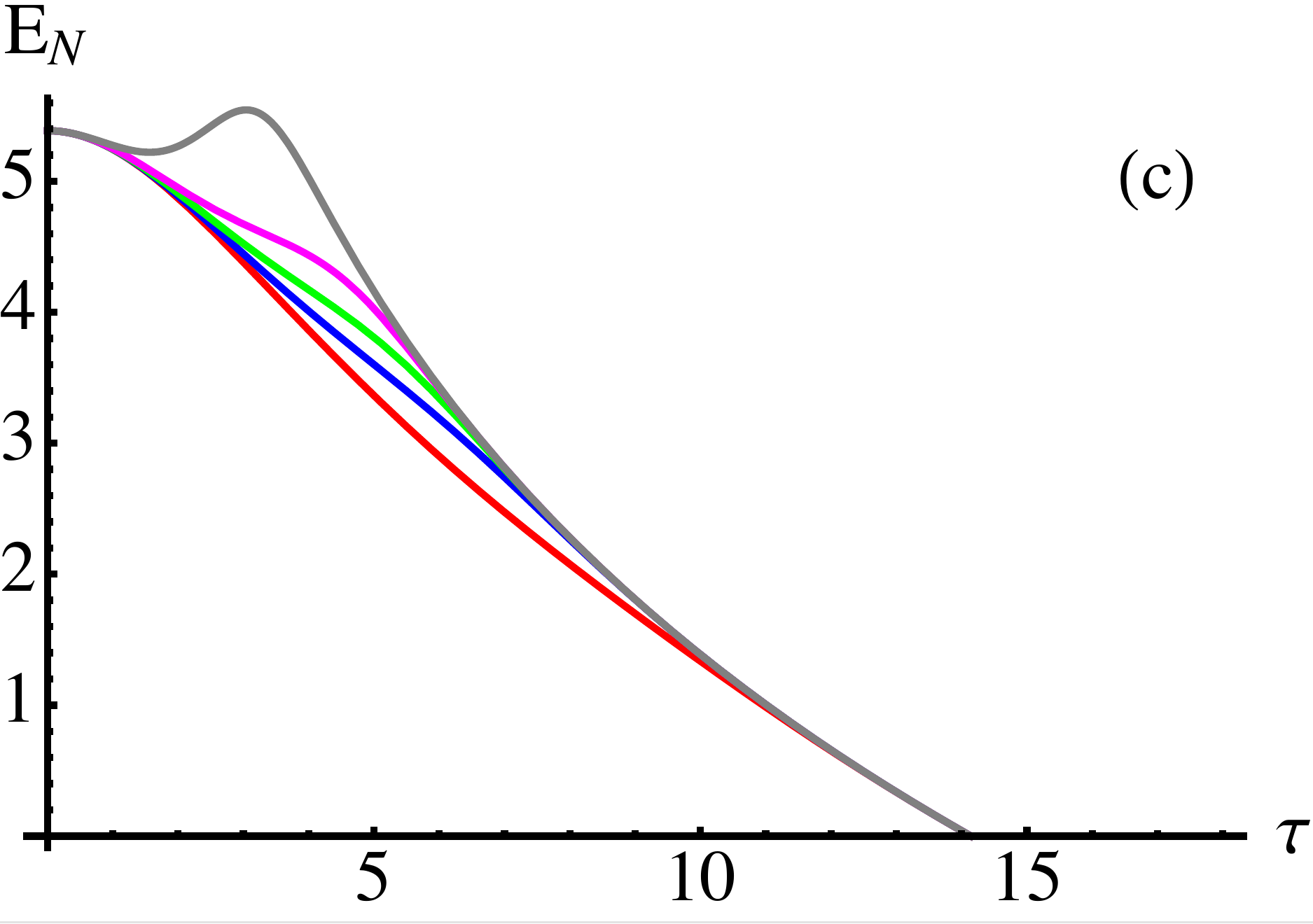}
\caption{\corr{Entanglement negativity $E_N$ as a function of the \corrbina{dimensionless} time $\tau$ for different values of the involved \corrbina{(dimensionless)} parameters. In panel (a) we show results for $\Omega=1$, $J_0\delta=0.01$ and different values of $r$, from top to bottom $r=10, 2, 1, 0.5, 0.1$, corresponding to gray, magenta, green, blue and red curve, respectively. In panel (b) we show results for $\Omega=1$, $r=1$, and
different values of the product $J_0\delta$,
from top to bottom $J_0\delta=10^{-3}, 10^{-5/2}, 10^{-2} , 10^{-3/2}, 10^{-1}$ [same colors as in panel (a)]. 
In panel (c) we show results for $J_0\delta=0.01$, $r=1$, and different 
values of $\Omega$, from top to bottom $\Omega=10, 2, 1, 0.5, 0.1$ }[same colors as in panel (a)]. 
\label{f2}}
\end{figure}

Having established that the secular approximation holds in a rather wide of the parameter range and for a rather long period of time, we now proceed by studying the dynamics of entanglement, i.e. \corr{we study the behaviour of $E_N$ [calculated according to Eqs. (\ref{EN}) and (\ref{kappa})] as a function of time for different values of the involved parameters. Results are reported in Fig. \ref{f2}, where we show $E_N$ as funtion of $\tau$ for: $\Omega=1$, $J_0\delta=0.01$ and different values of $r$ [panel (a)],  
$\Omega=1$, $r=1$, and
different values of the product $J_0\delta$ [panel (b)], $J_0\delta=0.01$, $r=1$, and
different values of $\Omega$ [panel (c)].}
%%%
\par
The first observation is that the $E_N$ is not monotone in time, a clear signature of non-Markovianity \cite{Maniscalco07,cazza13}. \corr{Moreover, we observe the phenomenon of entanglement sudden death \cite{YuLett, YuScience} which occurs at a time  determined only by the bandwidth [compare results in panels (a) and (c), where $\delta$ is fixed, to those in panel (b) where $\delta$ is varying].

Overall, the results of Fig. \ref{f2} may be summarised as follows. The initial entanglement (determined by $r$) influences the dynamics at short times, but then the behaviour becomes universal and determined by the properties of the environment. In particular, sudden death occurs at a time determined by the bandwidth. The location of the bandwidth influences entanglement to a lesser extent. On the other hand, it may be fruitfully exploited, since increasing $\Omega$ leads to an increase of non-Markovianity and, in turn, to a temporary increase of entanglement
[see panel (c)], at least for large $\Omega$.
}
\par
\section{Conclusions} \label{s:concl} 
In this work, we have analyzed the entanglement dynamics in optical media 
characterized by a finite bandwidth. Upon assuming weak coupling and low temperature, 
we have obtained an exact analytic solution for the time dependent two-mode covariance 
matrix describing a Gaussian state of our system in the short time non-Markovian limit. 
Our results show that attenuation (damping) is strongly suppressed whereas 
the diffusion term depends only on the bandwidth.
\par
We have investigated the entanglement dynamics as a function of the bandwidth, 
the natural frequency and the initial amount of entanglement and show that there exist a wide range of situations where
decoherence is not much detrimental and entanglement may persist for a longer time. 
We have also proved that secular terms may be neglected in the short time 
non-Markovian limit. Our results are encouraging and show that materials with 
a photonic bandgap may provide a reliable way to transmit entanglement over long distance.
%%%%%%%%%%%%%%%%%%%%%%%%%%%%%%%
\section*{Acknowledgments}
\noindent This work has been supported by Khalifa University through project no. 8474000358 (FSU-2021-018). This work is devoted to the memory of Margherita. 
%%%%%%%%%%%%%%%%%%%%%%%%%%%%%%%

\end{document}